\documentclass{emulateapj}
\usepackage{epsf}
\usepackage{apjfonts}

\newcommand{\vla}{VLA\,1623}

\newcommand{\HCOP}{HCO$^+$}
\newcommand{\ntwoh}{N$_2$H$^+$}
\newcommand{\HICOP}{H$^{13}$CO$^+$}

\newcommand{\jone}{J=1$\rightarrow$0}

\newcommand{\jthree}{J=3$\rightarrow$2}
\newcommand{\jfour}{J=4$\rightarrow$3}

\newcommand{\kms}{km\,s$^{-1}$}

\slugcomment{To appear in ApJ 20 August 2006, v647n 2 issue}

\shorttitle{Kinematics in $\rho$ Oph A}
\shortauthors{Narayanan \& Logan}

\begin{document}
\title{Kinematics of Protostellar Objects in the $\rho$ Ophiuchus A
Region} 
\author{Gopal Narayanan and Daniel W. Logan} \affil{Department of
  Astronomy, University of Massachusetts, Amherst, MA 01003}
\email{gopal@astro.umass.edu, logan@nova.astro.umass.edu}

\begin{abstract}
We present the detection of infall, rotation and outflow kinematic
signatures towards both a protostellar source, \vla\ and what was
initially thought to be a pre-protostellar core, SM1N, in the $\rho$
Ophiuchus A region.  The kinematic signatures of early star formation
were detected in the dense molecular gas surrounding the embedded
sources using high signal-to-noise millimeter and submillimeter data.
Centroid velocity maps made with \HCOP\ \jfour\ and \jone\ line
emission exhibit the blue bulge signature of infall, which is
predicted to be seen when infall motion dominates over rotational
motion.  Further evidence for infalling gas is found in the \HCOP\
blue asymmetric line profiles and red asymmetric opacity profiles.  We
also performed CO \jthree\ and \jone\ observations to determine the
direction, orientation, and extent of molecular outflows, and report
the discovery of a new bipolar outflow possibly driven by SM1N.

\end{abstract}


\section{Introduction}

Observationally, the earliest stage of evolution for low-mass
protostars is the so-called ``Class 0'' phase, which is characterized
by a blackbody-like spectral energy distribution (SED) that peaks in
the submillimeter (\citet{awb93}, hereafter AWB93). Barsony (1994)
defines the following as characteristics of Class 0 objects:
$L_{bol}/L_{1.3 mm} \leq 2\times 10^4$, an SED like a 30~K blackbody,
undetected at $\lambda < 10 \mu$m, and possession of a molecular
outflow. These characteristics of Class 0 objects imply that they are
in a very active mass accretion phase, and possess more circumstellar
mass than stellar mass. An even earlier phase of evolution prior to
class 0 objects has also been proposed, the so-called
``pre-protostellar'' stage \citep{ward94}. Observationally, objects in
this stage of evolution have SEDs that are similar to Class 0 objects,
but they lack other kinematic signatures of star formation such as
molecular outflows and centimeter continuum emission. The
pre-protostellar cores are probably gravitationally bound, but lack a
central hydrostatic stellar source, and are hence not yet
``protostars''. Indeed molecular line spectroscopy of a number of
pre-protostellar cores (e.g. \citet{taf98}) show that they are
undergoing infall, but more slowly than that seen towards Class 0
protostars.

To study the detailed kinematical evolution of pre-protostellar
objects to Class 0 objects and beyond, it is important to study both
type of objects with observational tools like millimeter and
submillimeter spectroscopy. Pre-protostellar cores do not possess a
strong outflow, and hence it is expected to be less complex to
disentangle the various motions that accompany star formation such as
infall, rotation, outflow and turbulence.  Since the details of the
star formation process can vary from cloud to cloud, and may depend on
environmental factors, it is crucial to choose pre-protostellar and
protostellar candidates within the same region. The centrally
condensed L1688 core of the nearby (160 pc) $\rho$ Ophiuchus
(hereafter $\rho$ Oph) molecular cloud consists of a rich array of
young stellar objects at various stages of evolution
\citep{wilking83}. In particular, the cloud A region (called $\rho$
Oph A) harbors a number of pre-protostellar and protostellar cores.
In this region, there is \vla, which was designated as the prototype
of the Class 0 phase of evolution (AWB93).  \vla\ is known to possess
a remarkably collimated and very young ($\sim 6000$ yrs) molecular
outflow (\citet{amd90}, AWB93). The millimeter and submillimeter
continuum maps of AWB93 reveal that $\rho$ Oph A breaks up into 4
cores, SM1, \vla, SM1N, and SM2. Of these, \vla\ was the only source
with an identified outflow. Both SM1N and SM1 are probably real
pre-protostellar cores (AWB93). Wide-field 1.3 mm continuum mapping by
\citep{motte98} (hereafter MAN98) reveals that Oph A is the brightest
of all Oph cores, and in addition to \vla\ there are eight other
starless cores in $\rho$ Oph A. The objects A-MM6, A-MM7, and A-MM8,
together with SM1N, SM1 and SM2, compose an arc within Oph A, which
seems to be delineating a photo-disassociation front of a PDR region
illuminated by the nearby B star S1 (see MAN98). Recent high angular
resolution N$_2$H$^+$ observations \citep{difrancesco04} of the
quiescent dense gas in this region show emission peaks of this tracer
at SM1 and SM1N, but not at \vla. N$_2$H$^+$ can deplete significantly
at densities greater than $10^5$ cm$^{-3}$ \citep{bergin02}, so the
difference in N$_2$H$^+$ emission from SM1N to \vla\ might be another
indicator that SM1N is less evolved than \vla.

In this paper, we present results of a study of the $\rho$ Oph A
region designed to make a comparison of pre-protostellar core(s) with
a more evolved Class 0 protostellar object (\vla).  One of the main
goals of this study was to understand better the earliest stages of
star formation from the pre-protostellar phase to the Class 0
phase. However, as will be seen from the results of this study, SM1N
seems to power an outflow, and is hence probably not
pre-protostellar. To probe the kinematics and physical conditions of
the cores, we performed a multi-transitional study using both density
sensitive molecular tracers, and CO and its isotopes. \HCOP\ and
\ntwoh\ preferentially trace high density gas, while CO traces a much
wider range of gas densities, in addition to being a good tracer of
outflows. Submillimeter transitions of \HCOP\ and CO probe regions of
higher density and excitation than the millimeter transitions. For
this reason, we augment our millimeter observations with submillimeter
observations of corresponding transitions. In \S2, we describe our
observations, and in \S3 we present our results.

\section{Observations}
\subsection{FCRAO}

Millimeter observations of \HCOP, \HICOP and \ntwoh\ towards \vla\ and
SM1N were performed with the Five College Radio Astronomy Observatory
(FCRAO) 14m telescope in December 1998, February 1999, and June 2001
using the SEQUOIA 16-beam array receiver \citep{ege99}, and the FAAS
backend of 16 autocorrelation spectrometers.  Mapping of the $^{12}$CO
and $^{13}$CO J=1$\rightarrow$0 transitions were also performed in
February 2004 using the newly-developed on-the-fly (OTF) mapping
technique at FCRAO using the upgraded SEQUOIA array receiver which had
dual-polarization capability, and a dual channel correlator (DCC)
backend that allowed the observation of both the $^{12}$CO and
$^{13}$CO isotopes simultaneously with a spectral resolution of 50
kHz. The 2004 observations covered a $12^\prime \times 12^\prime$ area
of the $\rho$ Oph A cloud. Table~\ref{observations} summarizes the
frequencies and effective resolutions of each transition.  \HCOP\ and
\ntwoh\ line transitions were observed via frequency-switching with
subsequent folding and third-order baseline removal.  \HCOP\ and its
isotopic observations span a region $6' \times 6'$ centered on \vla\
($\alpha (1950) = 16^h23^m24.8^s, \delta (1950) =
-24^o17'46'')$. Pre-OTF data were reduced and analyzed with the Gildas
CLASS software package. The OTF data was reduced using the FCRAO
OTFTOOL software package \citep{heyer2004}.  Pointing and focus were
verified with observations of SiO masers.

\subsection{CSO}

We obtained sub-millimeter observations using the 10.4~m telescope of
the Caltech Submillimeter Observatory (CSO) at Mauna Kea, Hawaii.  The
May 1996 \HCOP\ and CO observations were performed using On-The-Fly
(OTF) mapping.  These observations were made with the 345 GHz SIS
waveguide receiver \citep{ess89} and a 1000 channel, 50 MHz wide
acousto-optical spectrometer.  \HCOP\ observations span a region 70\arcsec\
$\times$ 70\arcsec\ centered on both \vla\ and SM1N.  Our CO observations
mapped a region of 150\arcsec $\times$ 150\arcsec\ centered on \vla\ and
SM1N. Individual OTF maps made on \vla\ and SM1N were combined
together to derive the results presented below. Data were reduced and
analyzed using the CLASS software package.

\begin{deluxetable}{ccccc}[!hb]
\tablewidth{0pt}
\tablecaption{Observations \label{observations}}
\tablehead{
\colhead{Molecule} & \colhead{Transition} & \colhead{Observatory} & \colhead{Frequency (GHz)} & \colhead{Beam Width (``)}}
\startdata
\HICOP\ & J = 1$\rightarrow$0 & FCRAO & 86.754288 & 62\\
\HCOP\ & J = 1$\rightarrow$0 & FCRAO & 89.188496 & 60\\
N$_2$H$^+$ &  123$\rightarrow$012 & FCRAO & 93.173770 & 58\\
$^{13}$CO & J = 1$\rightarrow$0 & FCRAO & 110.201354 & 49\\
CO & J = 1$\rightarrow$0 & FCRAO & 115.271202 & 47\\
$^{13}$CO & J = 3$\rightarrow$2 & CSO & 330.587980 & 23\\
C$^{17}$O & J = 3$\rightarrow$2 & CSO & 337.061123 & 22\\
CO & J = 3$\rightarrow$2 & CSO & 345.795990 & 22\\
\HICOP\ & J = 4$\rightarrow$3 & CSO & 346.998338 & 21\\
\HCOP\ & J = 4$\rightarrow$3 & CSO & 356.734134 & 21\\
\enddata
\end{deluxetable}

\section{Results}

\subsection{Molecular Line Profiles}

\begin{figure}[!htb]
\epsscale{1.1}
\plotone{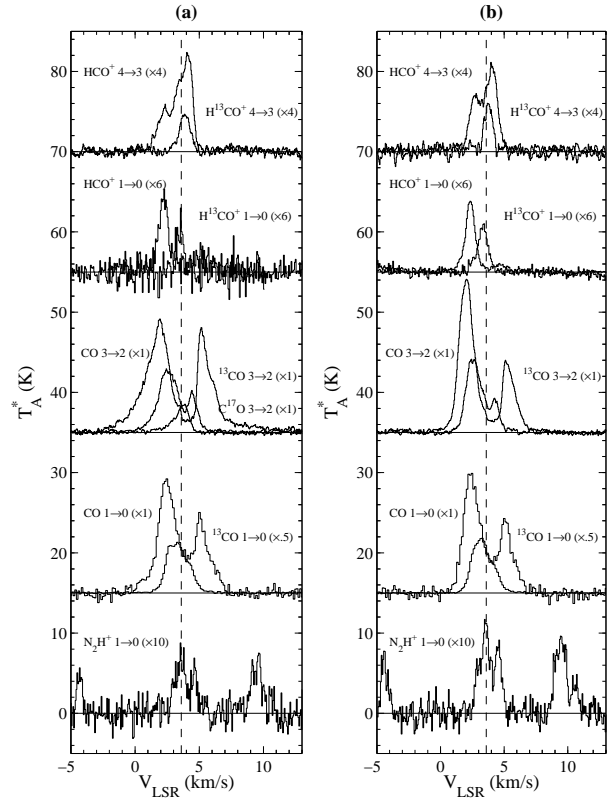}
\caption{Observed line profiles toward the central positions of (a)
\vla\ (left) and (b) SM1N (right).  The multiplication factor is
stated after the names of each line.  Both velocity ranges are -5 to
13 \kms.  The v$_{LSR}$ of $\sim 3.6$ \kms\ obtained using \ntwoh\ is
shown by the vertical dotted lines. The systemic velocity of the cloud
is $\sim 4$ \kms. \label{spectra}}
\end{figure}

Figure~\ref{spectra} shows a summary of the spectral observations
towards the central positions of two sources, \vla\ and SM1N.  The
left panel shows the observations towards \vla\ and the right panel
shows the observations towards the source SM1N. Table~\ref{tabvc}
lists the derived centroid velocities for various molecular
transitions towards the central position of \vla\ and SM1N. In
order to exclude any effects of outflows, the centroid velocities in
Table~\ref{tabvc} were computed within a velocity interval of $2$ to
$6$ \kms. 

\begin{deluxetable}{cll}[!htb]
\tablewidth{0pt}
\tablecaption{Centroid Velocities of Central Line Profiles \label{tabvc}}
\tablehead{
	\colhead{Source} & \colhead{Transition} & \colhead{$v_c$ (\kms)}\\ 
	\colhead{$v_{LSR}$ (\kms)} & \colhead{} &
	\colhead{}}
\startdata

VLA 1623 & CO \jone\ & $3.70 \pm 0.02$ \\
(3.6\tablenotemark{a} ) & $^{13}$CO \jone\ & $3.36 \pm 0.01$\\
& CO \jthree\ & $4.01 \pm 0.01$\\
& $^{13}$CO \jthree\ & $3.32 \pm 0.03$ \\
& C$^{17}$O \jthree\ & $3.53 \pm 0.06$ \\
& \HCOP\ \jone\ & $3.06 \pm 0.36$ \\
& \HICOP\ \jone\ & $3.91 \pm 0.11$\\
& \HCOP\ \jfour\ & $3.56 \pm 0.03$\\
& \HICOP\ \jfour\ & $3.81 \pm 0.06$\\

\hline



SM1N & CO \jone\ & $3.60 \pm 0.02$\\
(3.6\tablenotemark{a} ) & $^{13}$CO J=1$\rightarrow$0 & $3.32 \pm 0.01$\\
& CO \jthree\ & $3.60 \pm 0.01$\\
& $^{13}$CO \jthree\ & $3.16 \pm 0.04$ \\
& \HCOP\ \jone\ & $2.9 \pm 0.4$ \\
& \HICOP\ \jone\ & $3.34 \pm 0.16$ \\
& \HCOP\ \jfour\ & $3.53 \pm 0.04$\\
& \HICOP\ \jfour\ & $3.75 \pm 0.09$\\
\enddata
\tablenotetext{a}{$v_{LSR}$ derived from $N_2H^+$ hyperfine structure}
\end{deluxetable}
Since \ntwoh\ is typically optically thin, and is a good tracer of
dense, quiescent gas (\citet{benson98} and references therein), we
used the \jone\ observations of \ntwoh\ to determine the true systemic
velocity towards these sources. Hyperfine fitting of the \ntwoh\ lines
\citep[c.f.][]{caselli95} yields a v$_{LSR}$ of $3.63\pm0.02$
and $3.58\pm0.01$ \kms\ towards \vla\ and SM1N respectively.

The central spectra of \HCOP\ and CO towards \vla\ and SM1N show
strong self-reversed dips. These dips imply true self-absorption as
the optically thin tracers (\ntwoh, C$^{17}$O, \HICOP) peak at the
velocity of the self-reversal. 
One interesting feature of the submillimeter CSO \HCOP\ \jfour\
spectra is the lack of the ``classic'' blue asymmetry expected in
infalling regions.  It is this ``classic'' blue asymmetric line
profile signature that has been the technique of choice for most
studies of infall (see for example, \citet{gez97}, hereafter
GEZC97). Indeed, the lack of blue asymmetry in their \HCOP\
observations led GEZC97 to conclude that \vla\ was an ambiguous case
for the interpretation of infall. In fact, it can be seen from
Figure~\ref{spectra} that the central \HCOP~ \jfour\ line profiles in
both sources exhibit a red asymmetry, a profile consistent with
expanding rather than infalling gas. Indeed, a very strong and
collimated outflow has been discovered towards \vla\ \citep{amd90}. It
is likely that the expected blue-asymmetric line profile signature of
infall has been inverted in the presence of this strong outflow.  The
isotopic \HICOP~ \jfour\ has a centroid velocity (see
Table~\ref{tabvc}) which is red-shifted with respect to the main line
(as might be expected for infall), but at the same time is slightly
more red-shifted with respect to the $\sim 3.6$\kms\ velocity derived
from the \ntwoh\ transition. In the presence of pure infall, the
optically thin \HICOP\ line would be expected to be centered on the
systemic velocity of the cloud core. As will be shown in
\S\ref{impact}, \HCOP\ and its isotopes do trace the outflows, so this
may cause the velocity centroid of these lines to move systematically
redward, compared to the pure infall scenario.

In comparison with their corresponding \HCOP~ \jfour\ transitions, the
\HCOP~ \jone\ spectra are heavily blue shifted. Their self-absorption
dips almost reach the continuum baseline, and the red-shifted peak is
barely visible within the signal-to-noise ratio of our observations.
Why are the \HCOP~ \jone\ observations so different in appearance from
the \HCOP~ \jfour\ observations?  The two transitions are likely
probing differing regions of gas.  The gas observed by the larger beam
of the FCRAO 14~m in \HCOP~ \jone\ comes from extended regions around
the two sources.  The extended envelope traced by the lower transition
may be indicative of the overall large scale motion of the gas due to
collapse.  The CSO \HCOP~ \jfour\ transition probes a smaller region
of gas as well as probing farther into the envelope.  Thus, the
shorter wavelength submillimeter transition may be picking out the
densest gas regions near the protostellar cores.  Probing nearer the
source allows detection of the inner region of infall but also the
developing outflow, which might cause the red asymmetry seen in the
submillimeter transitions.

The dipole moment of CO is much smaller than \HCOP~ and is thus
expected to trace a wider range of densities. CO is often seen tracing
swept-up and outflowing material, whereas the \HCOP\ isotopes probe
more of the denser, centrally condensed gas.  Indeed the broad wings
of emission seen for the CO \jthree~ spectra in Figure~\ref{spectra}
imply that CO is tracing outflows in both sources. Curiously, the
central line profiles of CO also show the ``classic'' blue asymmetry
already discussed.  The CO spectra show deep self absorption in the
line core and very broad wings of emission that extend beyond $\pm 10$
\kms\ from the systemic velocity.  The self-absorption dip and blue
asymmetry profile of the $^{13}$CO \jthree\ support the
notion that this isotope is optically thick as well.  The rarer
isotope C$^{17}$O has a gaussian line profile which peaks at the
v$_{LSR}$ of \vla, as expected for an optically thin line.


In the presence of infall, the line emission from the rarer isotope is
expected to be red-shifted with respect to that from the main isotope
\citep{nwb98, nar02}.  In the cases of VLA 1623 and SM1N, all of the
centroid velocities of the rarer isotopic spectra are red-shifted with
respect to their main lines (see Table~\ref{tabvc}).  If infall were
the only source of motion, both the emergent line intensity and
optical depth profile must be asymmetric, with the line profile
showing blue asymmetry, and the optical depth profile showing red
asymmetry \citep{nwb98}.  In the presence of pure expansion (outflow),
the converse would be true: the line profile would be red asymmetric,
and the optical depth profile would be blue asymmetric.

\begin{figure}[!htb]
\epsscale{1.2}
\plotone{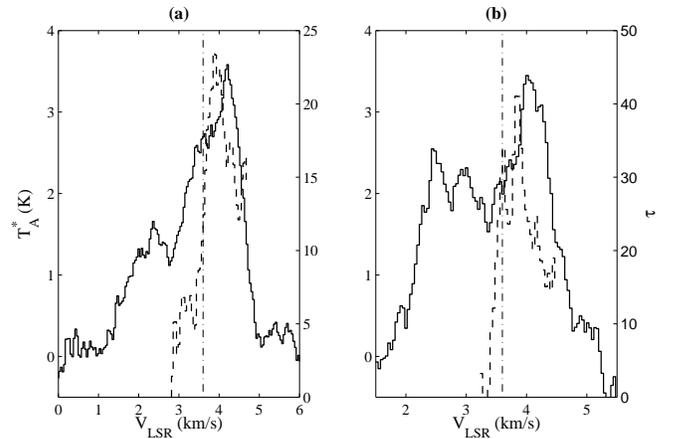}
\caption{\HCOP\ \jfour\ optical depth profiles toward the center
positions of (a) VLA1623 (left) and (b) SM1N (right).  The temperature
scale is shown on the left-hand side of the Y axis, while the opacity
scale shown on the right-hand side Y axis.  The velocity for the
ambient cloud (3.6 \kms) derived from our \ntwoh\ observations is
shown as the vertical lines.  The line profile of \HCOP\ is shown in
the solid red histograms and the opacity profile is shown as the
dashed blue histograms and only over the velocity range of the
\HICOP\ line profile. \label{opacities}}
\end{figure}

We performed an optical depth analysis using the CSO HCO$^+$ and
H$^{13}$CO$^+$ \jfour\ observations.  In Figure~\ref{opacities}, we 
present the optical depth profiles for VLA 1623
and SM1N.  The optical depth calculations were performed only over the
FWHM of the isotopic lines.  The opacities were estimated from
the observed line profile ratios of \HCOP/\HICOP.  The opacity,
$\tau_\nu^M$ of the more abundant isotope was estimated using
I(Main)$/$I(Iso)$ = (1-e^{\tau_\nu^M})/(1-e^{\tau_\nu^{M/r}})$, where
$r$ is the ratio of main to isotopic abundances, and I(Main) and
I(Iso) are the line intensities of the main and isotopic lines
respectively.  An isotopic abundance ratio of [\HCOP]:[\HICOP]$=$ 45:1
was used. This technique assumes that the excitation temperature is
the same for the main and isotopic species.

From Figure~\ref{opacities}, we see that not only are the main line
profiles red asymmetric towards both \vla\ and SM1N, but the optical
depth profiles also exhibit a red asymmetry.  The red-shifted line
profiles are consistent with expansion or outflow but the red-shifted
opacity profiles are consistent with tracing infall. The main isotope
\HCOP~ \jfour\ must be tracing a significant part of the outflow for
its line profile to show an expansion signature. However, the opacity
profile which is principally determined by the more optically thin
isotope seems to be more consistent with tracing infall. The optically
thin isotope is likely tracing material further towards the center of
the cloud cores, whereas the main line is likely being contaminated by
the expanding (dense) shells of the outflow. This indicates the
importance of using both isotopes in the interpretation, and that the
submillimeter transitions of \HCOP\ and its isotope are tracing both
outflowing material and the material that is participating in
gravitational collapse near the cores towards both VLA 1623 and SM1N.


\begin{figure*}[!htb]
\epsscale{1.0}
\centering
\includegraphics[angle=-90,scale=0.65]{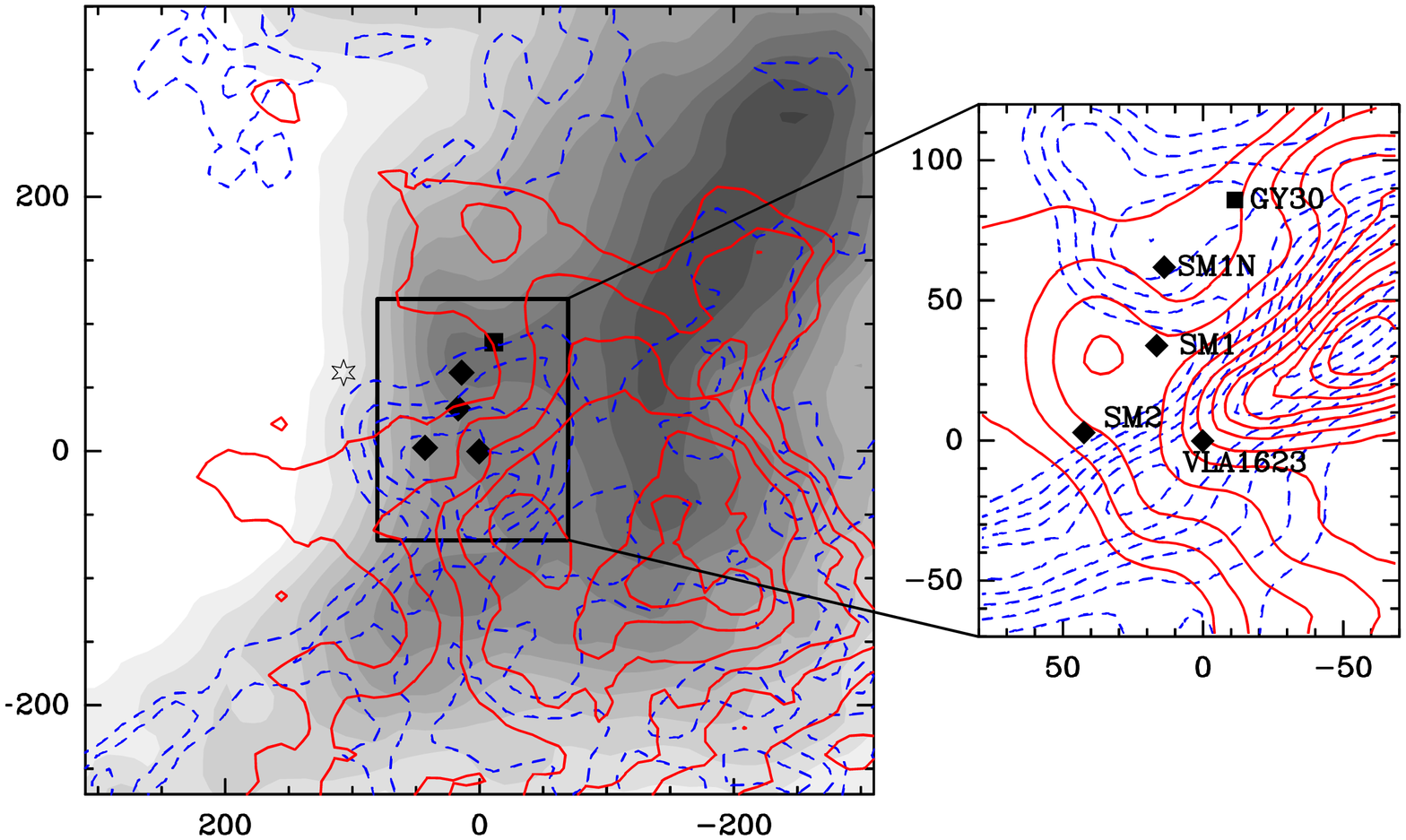}
\caption{Outflows in the region. The left panel shows the $^{13}$CO
\jone\ line-core emission (in grey-scale) and the $^{12}$CO \jone\
line-wing emission (in contours) observed with the FCRAO 14~m. The
$^{13}$CO emission is integrated over 1 to 6 \kms, and the $^{12}$CO
over -3 to 2 \kms (blue shifted -- dashed contours), and 5 to 10 \kms
(red shifted -- solid contours). The contour levels in K$\cdot$\kms\
for the \jone\ data on the left are as follows: 15 to 35 by 1.5
($^{13}$CO), 6 to 25 by 1.5 ($^{12}$CO blue), 6 to 26 by 2 ($^{12}$CO
red). The inset to the right shows the inner $190^{\prime\prime}\times
150^{\prime\prime}$ region mapped in $^{12}$CO \jthree\ emission. Blue
shifted emission (-8 to 2 \kms; dashed lines) has contour levels from
8 to 30 by 2 K$\cdot$\kms\ while red shifted emission (5 to 15 \kms)
has contour levels from 8 to 30 by 2 K$\cdot$\kms. The star on the
left panel shows the location of the nearby B star S1 (see MAN98),
while the diamonds in both panels show the location of the embedded
submillimeter continuum objects. The square shows the location of the
2.111$\mu$m near-infrared source GY30 (see KSHK). The X and Y axes of
both plots show the angular offsets in arc-seconds from the position of
\vla. \label{hvco}}
\end{figure*}

\subsection{Molecular Line Maps}
\subsubsection{Outflows}

Outflows play an important role in the early phases of stellar
evolution.  In fact, bipolar outflows in Class 0 sources, like VLA
1623, indicate that outflows arise in star formation at a very early
stage.  The collimated outflow observed toward VLA 1623 has been
studied extensively since its discovery by \citet{amd90}.


Figure~\ref{hvco} presents the newly obtained millimeter and
submillimeter CO maps towards the $\rho$ Oph A region. The left panel
shows the FCRAO CO maps in the \jone\ transition. Curiously, the
grey-scale $^{13}$CO image shows a ridge of emission that peaks about
3\arcmin\ west of the submillimeter continuum sources, SM1, SM1N, VLA
1623 and SM2. The $^{13}$CO map also shows a well-defined edge of
emission to the east close to the location of the nearby B star S1. It
has been suggested that the SM1N, SM1 and SM2 cores form an arc of
filamentary structures centered on the B star S1, possibly indicating
a link between these submillimeter clumps and star S1 (MAN98,
\citet{difrancesco04}). The left panel of Figure~\ref{hvco} also shows
the red-shifted and blue-shifted line wing emission in $^{12}$CO
\jone\ emission. Centered on VLA 1623, we can see the outflow,
especially the more collimated blue-shifted lobe of the outflow. The
red-shifted line wing emission appears more diffuse throughout the
region without a clear signature representing bipolarity of the
outflow. A second blue-shifted component is seen east and north of VLA
1623.

The submillimeter CSO CO \jthree\ data have higher angular resolution
and in this higher transition, excitation conditions are more
favorable for picking up the strong line wing emission from
outflows. The right panel of Figure~\ref{hvco} shows the integrated
intensity maps from the CO \jthree\ line wing emission over a smaller
region in $\rho$ Oph A. Indeed, in this transition the highly
collimated outflow of VLA 1623 is readily seen extending across the
map from the northwest to the southeast.  The superposition of the red
and blue-shifted emission indicates the VLA 1623 flow is located
almost in the plane of the sky.  Also, the differing angles ($\approx
20$\arcdeg) of the northwest red and blue outflow lobes is consistent
with the CO J=2$\rightarrow$1 observations from \citet{amd90} and
\citet{awb93}.  The north-west blue-shifted high velocity lobe
attributed to VLA 1623 may be a combination of an outflow from VLA
1623 more in line with the VLA 1623 red-shifted outflow lobe and a
separate blue-shifted outflow originating from some source north and
east of VLA 1623. North of the VLA 1623 flow, we see two
well-separated red-shifted and blue-shifted lobes. The midpoint of the
line joining the peaks of red-shifted and blue-shifted emission of
this flow north of VLA 1623 falls very close to SM1N. We suggest that
we have detected a new bipolar outflow centered on the source SM1N.

A second outflow was reported to have been discovered north of the VLA
1623 flow at approximately the position reported here, based on
smaller angular extent millimeter interferometric measurements
\citep[hereafter KSHK]{kam01}, and JCMT CO \jthree\ observations
\citep[hereafter KSHUK]{kam03}. However, both of these studies covered
a region only 1\arcmin\ centered on SM1N in right ascension. Our maps
cover 2.5\arcmin\ around SM1N in RA. As a result of the smaller scale
of their observations, KSHK and KSHUK claim the detection of a new
northern flow (which they dubbed ``AN outflow'') with a driving source
that is a near-infrared star called GY30 (which is $\sim 35$\arcsec\
from SM1N). They also conclude that the AN outflow has a blue-shifted
lobe in the east, and a red-shifted lobe in the west, which is the same
as the orientation of the VLA 1623 flow. A comparison of Figure 2 of
KSHUK with Figure~\ref{hvco} of this paper indicates that their
blue-shifted lobe coincides with the northern blue-shifted lobe seen in
our work. But KSHUK do not observe the region where the eastern
red-shifted lobe is seen. As a result, they have misidentified the
source and orientation of the AN outflow. It appears from our present
work that a new bipolar outflow is present north of VLA 1623, with a
driving source, which is most likely SM1N, and whose orientation of
blue and red-shifted lobes in the sky is opposite to that of VLA 1623.

We have estimated the physical parameters of the newly discovered
outflow towards SM1N using the CSO CO \jthree\ data (see
Table~\ref{outflow-parameters}). The outflow from \vla\ is not fully
mapped here, so we do not attempt to repeat the estimates of physical
parameters of the \vla\ outflow \citep[see for e.g.][]{amd90,
awb93}. In order to estimate the masses of the outflowing gas, local
thermodynamic equilibrium (LTE) conditions are assumed with an
excitation temperature of T$_{ex}=40$ K \citep{amd90}. The total
$^{12}$CO gas column density towards each lobe of the SM1N outflow is
estimated by using the integrated brightness temperature under the
line for velocities of $-8$ to $2$ \kms\ and $5$ to $15$ \kms\
respectively for the blue and red-shifted lobes, and assuming that the
$^{12}$CO emission is optically thin. A fractional abundance of
$X$($^{12}$CO) = $1\times 10^{-4}$ and a distance of 160 pc to
$\rho$ Oph are used.

We then estimated dynamical timescale, momentum, kinetic energy, and
mechanical luminosity of the flow. The dynamical timescale, $t_d$ was
estimated from outflow length and velocity that was assumed to be
constant. The differences between the systemic velocity and the
central velocities of blueshifted (redshifted) maps were used as the
outflow velocities $V$ of blueshifted (redshifted) components. The
calculated values of the momentum $P = MV$, kinetic energy $E =
MV^2/2$, and mechanical luminosity $L_m = E/t_d$ are summarized in
Table~\ref{outflow-parameters}. These quantities are not corrected for
a possible inclination angle of the outflow.

\begin{deluxetable}{lcc}[!htb]
\tablewidth{0pt}
\tablecaption{Physical Parameters of SM1N Outflow (Based on CSO CO \jthree)\label{outflow-parameters}}
\tablehead{
	\colhead{Parameter} & \colhead{Blue} & \colhead{Red} }
\startdata
Velocity (\kms) & 6 & 5.5\\
Length ($\times 10^4$ AU) & 1.2 & 1.1 \\
Mass ($\times 10^{-3}$ M$_\odot$) & 5.5 & 3.5\\
Dynamical Timescale ($\times 10^3$ yrs) & 9.2 & 10\\
Momentum ($\times 10^{-2}$ M$_\odot$\kms) & 3.3 & 1.9\\
Kinetic Energy ($\times 10^{42}$ ergs) & 2.0 & 1.0\\
Mechanical Luminosity ($\times 10^{-3}$ L$_\odot$) & 1.7 & 0.8\\
\enddata
\end{deluxetable}

This new discovery of the SM1N outflow may indicate either this
protostar is more evolved than previously believed or the onset of
outflows occur almost simultaneously with the first indication of
infall.

\subsubsection{Impact of Outflows}
\label{impact}

\begin{figure}[!htb]
\epsscale{0.95}
\plotone{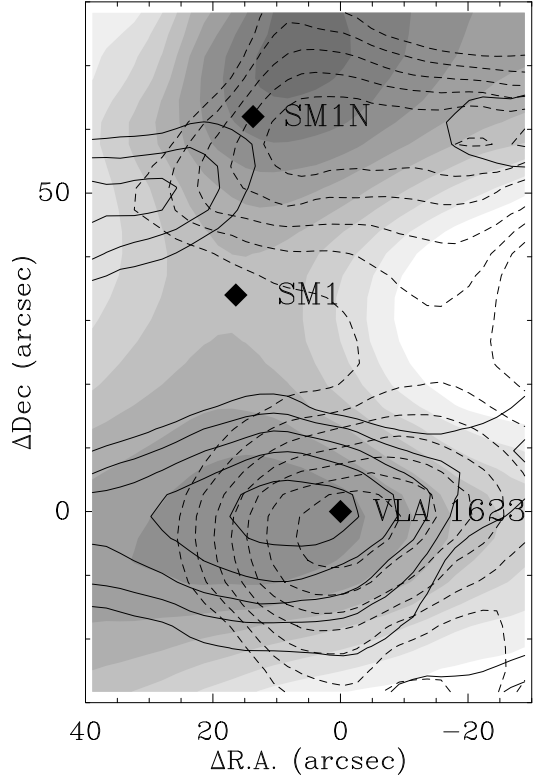}
\caption{CSO \HCOP\ \jfour\ Integrated Intensity Maps toward 
\vla\ and SM1N.  Blue-shifted emission (-2 to 2 \kms) is shown in dashed 
contours, line core emission (2 to 6 \kms) is shown in grayscale, and 
red-shifted emission (6 to 10 km/s) is shown in solid contours.  The 
contour levels are 0.1 to 0.45 by 0.05 K$\cdot$km/s for blue shifted 
emission, 2 to 6.5 by 0.5 K$\cdot$km/s for line core emission, and 0.1 
to 0.45 by 0.05 K$\cdot$km/s for red-shifted emission. \label{intensity}}
\end{figure}

With detected outflows in both sources, we can determine the extent to
which the \HCOP\ line profiles are affected by the molecular outflows
mapped in CO. To do this, we made integrated intensity maps of the
blue-shifted, line-core, and red-shifted emission of \HCOP\ \jfour.
Figure~\ref{intensity} shows the \HCOP\ \jfour\ integrated emission
around \vla\ and SM1N, where the line core integrated intensity is
shown in grayscale, blue-shifted emission in dashed contours, and the
red-shifted emission is shown in solid contours.  The line-core
emission is expected to trace the denser, infalling material, while
the line-wing maps may show contamination due to material swept-up by
the outflows.

Figure ~\ref{intensity} shows a complex morphology of both the line-core 
and line-wing emissions.  The line-core emission displays two orthogonal 
components, one elongated along the direction of the outflows and one 
perpendicular.  This effect is readily seen in the VLA 1623 region and to 
a lesser extent in the region around SM1N.  The overall structure of the 
line-core emission is consistent with the continuum map of the cloud core 
region in \citet{awb93}, Figure 1.

The line-wing emission around SM1N is collimated and consistent with
the high velocity CO map in Figure~\ref{hvco}.  The line-wing emission
of \HCOP\ confirms the detection of the outflow seen in CO \jthree\
emission. There are clear indications that the line-wing emission is
elongated along the outflow direction.


The HCO$^+$ line-wing integrated intensity about VLA 1623 is more
complex.  There is a definite elongation of red and blue-shifted
emission along the well known outflow driven by VLA 1623.  Although
there is a slight deviation at the edges of the map, the line-core,
blue-shifted, and red-shifted emission are spatially coincident along
the outflow.  In addition to the outflow elongation, the red-shifted
and blue-shifted emission has a small component perpendicular to the
outflow.  The red-shifted emission is predominately toward the
north-north-east and the blue-shifted emission is stronger in the
south-south-west direction.  This emission perpendicular to the
outflow direction may be tracing the overall rotation of the molecular
cloud about VLA 1623.

\subsubsection{Centroid Velocity Maps}
\label{centroid_map}

Given a velocity range, the centroid velocity is defined as that
velocity at which the integrated intensity is equal on both sides.  It
has been shown that in the presence of complicated velocity fields,
centroid velocity maps provide a better indication of the underlying
velocity fields than integrated intensity maps \citep{al88, wnb94,
nw98, nar02}. In the model isovelocity maps of \citet{nw98} which considered
a rotating infalling protostellar core, the rotational velocity field
imposes a gradient of blue-shifted to red-shifted velocities, with the
sense of the gradient being orthogonal to the rotational axis. When
infall dominates in the central regions, the line profiles in the
central region become blue asymmetric, and hence there is a
preponderance of blue-shifted velocities in the central regions of the
isovelocity maps, giving rise to the so-called ``blue-bulge''
signature. Here, we make centroid velocity maps from our data to see
if such an infall signature can be detected towards \vla~ and SM1N.

\begin{figure}[!htb]
\epsscale{0.95}
\plotone{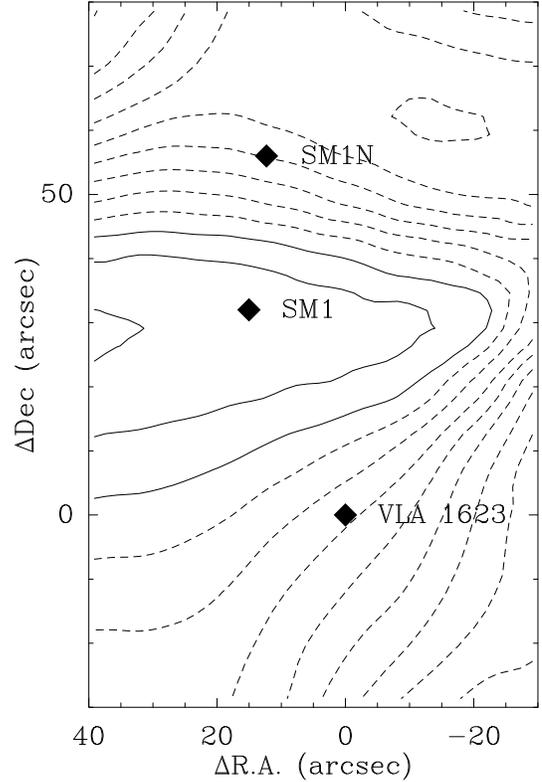}
\caption{CSO \HCOP\ \jfour\ centroid velocity map of VLA 1623 and
SM1N.  The centroid velocities only in the line core (2 to 6 \kms) are
shown.  The centroid velocities are compared with that of the
optically thin isotope ($\sim 3.8$ \kms). Blue-Shifted velocities (
$<$ 3.8 \kms) are shown in dashed contours, and red-shifted velocities
( $>$ 3.8 \kms) are shown in solid contours.  Contour levels are 3.0
to 4.3 \kms\ by 0.1 \kms. \label{centroid}}
\end{figure}

We present the \HCOP~ \jfour\ centroid velocity map of the \vla\ and
SM1N region in Figure~\ref{centroid}.  Blue-Shifted and red-shifted
emission are shown with dashed and solid isovelocity contours
respectively.  It has been shown in \S\ref{impact} that outflows show more of
an impact on the line-wings of \HCOP~ emission. In an effort to lessen
the impact of outflows on the centroid velocity map, the centroid
velocities are computed only over the line core emission (2 -- 6
\kms).

The isovelocity contours around VLA 1623 are parallel to the direction
of the outflow axis and show a gradient from the north-east to the
south-west.  The resulting gradient could be interpreted as a general
rotation of the gas around \vla\ with the south-western gas rotating
toward the observer and the north-eastern gas rotating away from the
observer.  Some evidence for this direction of rotation is also seen
in the \HCOP\ integrated intensity map of Figure ~\ref{intensity}.
The blue-shifted contours in \vla\ are even seen to extend north-east
of the axis of rotation.  The encroachment of the blue-shifted contours
across the rotation axis displays the predicted ``blue-bulge'' infall
signature.

The region around SM1N shows a more complex morphology in the centroid
velocity map.  If the gradient in isovelocity contours towards SM1N is
interpreted solely as that produced by rotation, one major feature is
that the sense of rotation is counter to that of \vla\ with the
blue-shifted gas north of the source and the red-shifted emission toward
the south.  Another interesting feature is that unlike in \vla\, the
isovelocity contours do not lie parallel to the outflow axis in
SM1N. Toward the eastern edge of the map, the contours curve to the
expected orientation if the contours were to align with the rotation
axis.  However, in the central regions, the gradient appears to be in
a north-north-west to south-south-east direction. Such a direction for
the gradient in \HCOP~ emission might result if even its line-core in
SM1N is tracing both outflow and infall, in which case the resultant
velocity gradient would be oriented at an angle intermediate between
the outflow direction and rotational gradient direction. Although
certainly less defined than the blue bulge found toward \vla, SM1N's
blue bulge is evident from the blue shifted shifted isovelocity
contours extending from the north-west through the source to the
south-east.

\section{Discussion}
\subsection{Infall Modelling}

To constrain the infall parameters in both sources, we modeled our
sources using a semi-analytic infall model based on the Terebey, Shu,
and Casen (TSC) collapse treatment \citep{tsc84}. For a more complete
explanation of the model used, see \citet{thesis, nwb98, nar02}.  

\begin{figure*}[!htb]
\epsscale{0.95}
\plottwo{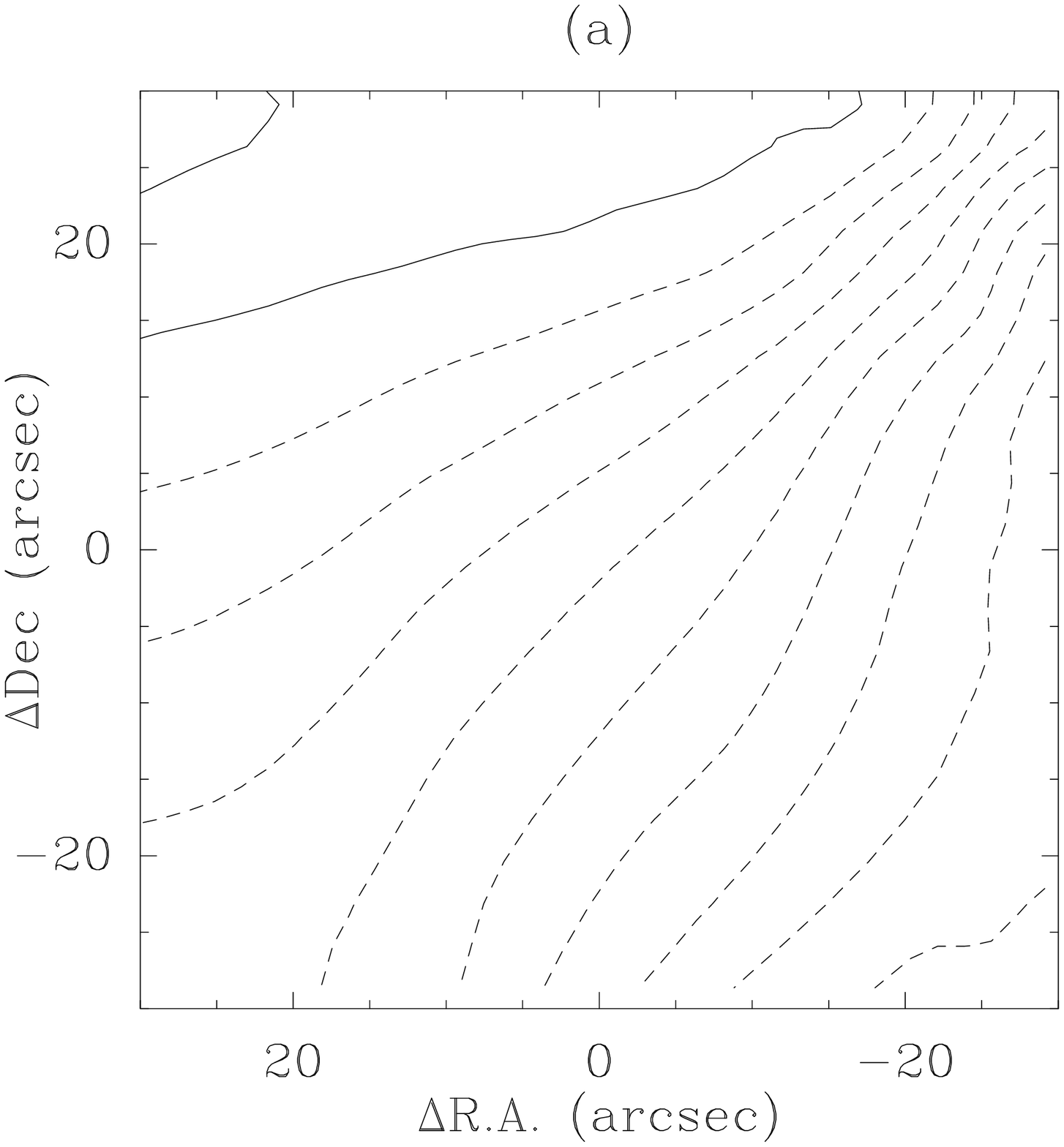}{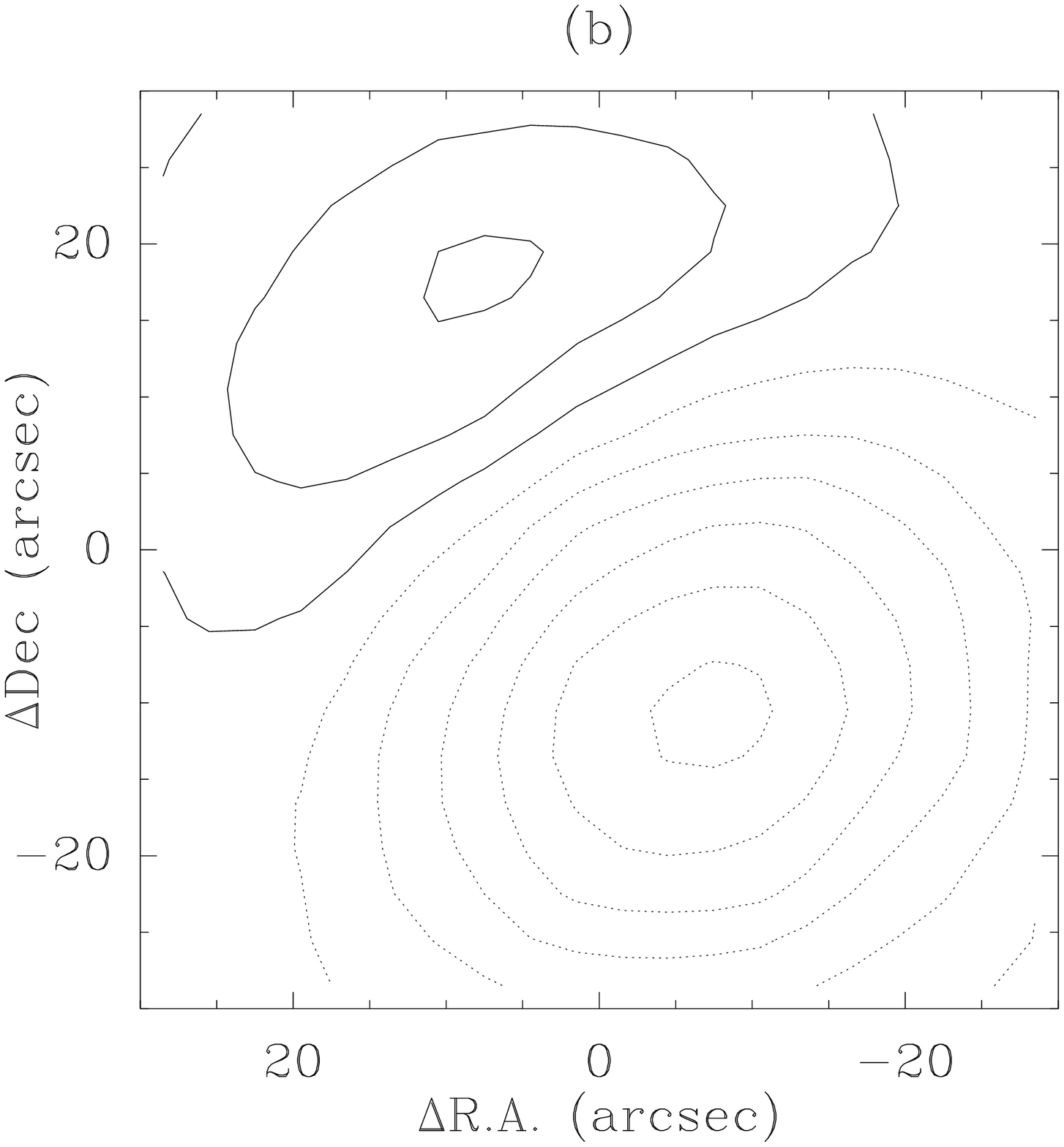}
\caption{ Centroid velocity maps towards \vla\ for (a) observed
  (left), and (b) modelled (right) in the \HCOP~ \jfour\ transition.
  The contour intervals are 2.8 to 4.5 by 0.1 \kms, and -.006 to 0.002
  by 0.001 \kms respectively.\label{modelvs}}
\end{figure*}

The TSC model requires the use of three parameters: the sound speed
$a$, the rate of cloud rotation $\Omega$, and the infall time $t$ (or
infall radius $r_{inf} = at$).  
After the model protostar has been generated, values for the
temperature, density, and velocity along the prescribed lines of sight
(LOSs) are generated for any given viewing angle.  The LOSs are
defined by two geometric angles: $\alpha$, the angle out of the plane of the
sky, and $\psi$, the angle in the plane of the sky.  After the program
generates all LOSs, a separate radiative transfer code calculates
molecular and continuum emission along all LOSs using the assumption
of local thermal equilibrium (LTE). Many model runs were made varying
the infall and geometric parameters each time. The synthetic
observations were then convolved to the resolution of the telescopes
used in the observations.  The best fit models were determined by eye
by constraining the model spectra to the observed \HCOP\ and \HICOP\
spectra.

\begin{figure}[!hb]
\includegraphics[height=3.3in,angle=-90]{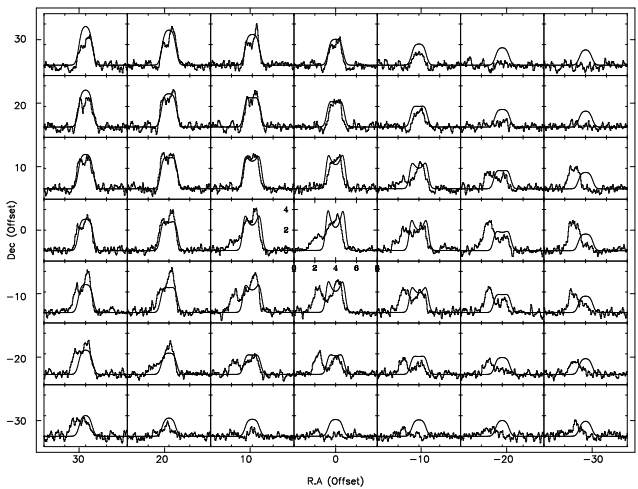}
\caption{TSC Model fits for VLA 1623.  Observed \HCOP\ \jfour\ 
(histograms) and model \HCOP\ \jfour\ (solid lines) spectral mosaics 
are shown. The velocity extent displayed is 0 to 8 \kms, and the 
temperature scale is -1 to 5 K. \label{modelspectra}}
\end{figure}

The TSC model does not include outflows, while \vla\ and SM1N both
have vigorous young outflow systems. However, as has been discussed in
\S\ref{impact}, while the line-wing emission of \HCOP\ seems to trace
the outflows, the line-core emission itself seems to be less affected
by the presence of outflows. Indeed, for \vla, the centroid velocity
map shown in Figure~\ref{centroid} shows predominantly the impact of
rotation and infall, both of which are well-treated by the TSC
formalism. In our infall modeling, we only use the line-core \HCOP\ to
constrain the infall parameters. Even in this light, the infall
parameters derived here should be used with some caution, as the
models are for single isolated protostars, and not for clusters of
protostars, and hence not rigorously applicable to the observations.

The comparison between the observed spectra and the simulated spectra
of the best-fit model towards \vla\ is shown in
Figure~\ref{modelspectra}. While the spectral shapes are reasonably
well-fit (at least in the central regions), the best fit TSC model
does not do a good job of matching the observed centroid velocity map
observed in \HCOP\ \jfour\ (see Figure~\ref{modelvs}).  Given the
caveats discussed above of applying the TSC model for this study, this
latter result is not surprising. Similar model plots were made for
SM1N, but are not shown here.

Since there is some evidence that even the line-core emission of \HCOP\
in SM1N is being impacted by the observed outflow (see
\S~\ref{centroid_map} and Figure~\ref{centroid}), it was expected that
it would be harder to fit the SM1N observations to the TSC model. The
velocity gradient in Figure~\ref{centroid} seems to imply a more or
less east-west rotational axis if the gradient is purely interpreted
as due to that of rotation. In view of the possible contamination from
the outflow, in our modeling, we forced the rotational axis for SM1N
to be parallel to that of \vla. Because of this, the model centroid
velocity map for SM1N cannot reproduce the direction of the gradients
seen in the observed map of Figure~\ref{centroid}. However, the
general appearance and width of the line profiles are well modeled
even in the region towards SM1N.

When the effects of outflows are separable from the observations of
density-sensitive molecules such as \HCOP\ and CS, the TSC model can
be used reasonably effectively to derive infall parameters (see the
case of IRAS 16293 and SMM4 in \citet{nwb98} and \citet{nar02}).  The
results of the TSC models clearly indicate the difficulty in modeling
complex regions that consist of both infalling and outflowing motions.
In fact, the inability to adequately account for the presence of
outflows, severely handicaps the modelling of spectra and centroid
velocity maps along the axes of the outflows.  Despite this
shortcoming and the complexity of the region, our models are robust
enough to deduce key properties of the regions around \vla\ and SM1N.
The TSC models confirm \vla 's outflow is within the plane of the sky
to $\pm 15\arcdeg$ as initially proposed by \citet{amd90}.  Second, our
modelling of \vla\ could not reproduce the observations unless a
temperature gradient was applied to the TSC code.  The best fit
temperature gradient decreases at an angle of 32\arcdeg\ west of south.
The nearby B3 star, S1, provides the best physical explanation for
this temperature gradient.  
Our best model parameters are summarized in Table~\ref{parameters}.

\begin{deluxetable}{cccccccc}[!h]
\tablewidth{0pt}
\tablecaption{Best Fit Model Parameters \label{parameters}}
\tablehead{
	\colhead{Source} & \colhead{r$_{inf}$} & \colhead{a} & \colhead{t$_{inf}$} & \colhead{$\Omega$} & \colhead{$\alpha$} & \colhead{$\psi$} & \colhead{v$_{turb}$} \\ & \colhead{(pc)} & \colhead{(\kms)} & \colhead{(yr)} & \colhead{(s$^{-1}$)} & \colhead{($^o$)} & \colhead{($^o$)} & \colhead{(\kms)} }
\startdata

VLA 1623 & 0.01 & 0.33 & 3.0 x 10$^4$ & 3 x 10$^{-14}$ & 0 & -60 & 0.5 \\
SM1N & 0.006 & 0.58 & 1.0 x 10$^4$ & 7 x 10$^{-14}$ & 0 & -60 & 0.6 \\
\enddata
\end{deluxetable}

\subsection{Timescales}
One of the more difficult challenges in star formation is the determination 
of age.  This difficulty arises from the lack of a definitive definition for 
the age of a protostar and inability to accurately measure key time 
dependent observables such as mass infall rate and velocity dispersion.  
One of the first estimates for the age of \vla\ was derived by \citet{amd90} 
by examining the dynamical timescale for \vla 's molecular outflow.  Their 
conclusion, a dynamical age of $\sim$3000 yr, is similar to the age derived 
by AWB93 ($\lesssim$6000 yr) using the mass infall rate.

Using our submillimeter CO \jthree\ data, the dynamical age of \vla\ is
found to be $\sim$4000 yr, in agreement with \citet{amd90}.  However,
our modelling of \vla\ suggests an older age for the time since infall
by approximately 10 times (see Table~\ref{parameters}).  Given errors
in the calculations of both the dynamical and modelled age estimates,
the similarity of the infall and dynamical timescales leads to the
conclusion that large outflows develop almost from the onset of
collapse.

\section{Conclusions}

We have observed two cores in the $\rho$ Oph A molecular cloud to
investigate the kinematic signatures of star formation.  We obtained
isotopic and main line CO \jthree, \jone, and \HCOP\ \jfour, \jone\
data of a known Class 0 source, \vla, and a core that was previously
identified as a pre-protostellar core, SM1N.  Using more spatially
extended and sensitive CO maps than those of KSHK and KSHUK, we
confirm the detection of the second molecular outflow north of
\vla. This second outflow is most likely generated by the protostar
SM1N, and not by the near-infrared star GY30 as suggested by KSHK and
KSHUK. SM1N appears to be in a more progressed evolutionary stage of
star formation than previously thought.

Millimeter observations toward the source positions show the expected
blue asymmetry.  However, the submillimeter transitions of \HCOP\
\jfour\ display a line profile signature of expanding rather than
infalling gas.  Most likely, in \vla\ and SM1N, the higher density,
centrally condensed gas kinematics are greatly affected by the
outflows emanating from both sources.  Examination of the opacity
profiles, however, provides support that, despite outflowing material,
the core is still undergoing gravitational collapse.

We performed centroid analysis of \vla\ and SM1N to confirm the
existence of the ``blue bulge'' signature found in other star forming
regions.  The ``blue bulge'' found around \vla\ is very well aligned
with the axis of rotation and is strong, confirming that \vla\ is a
true protostar.  The centroid velocity map near SM1N also exhibits
``blue bulge'' signature but it is not as well aligned about a
rotation axis as seen in \vla.


The detection of infall and outflow motions towards SM1N suggests that
both kinematic motions occur simultaneously from the earliest stages
of star-formation.



\acknowledgements

We thank the staff of the CSO for their support with the observations.
We also thank Chris Walker for help with the CSO observations. Ron
Snell and Jonathan Franklin helped collect some of the FCRAO CO
data. Research at the FCRAO is funded in part by the National Science
Foundation under grant AST 01-00793.

Facilities: \facility{FCRAO}, \facility{CSO}

\end{document}